\documentclass[doublecol]{epl2} 

\usepackage{physics}
\usepackage{bm}
\usepackage{chemarr}
\usepackage{amsmath}
\usepackage{color}
\usepackage{amssymb}
\usepackage[caption=false]{subfig}

\newcommand{\bea}{\begin{eqnarray}}
\newcommand{\eea}{\end{eqnarray}}

\newcommand{\blue}{\textcolor{black}}

\title{First Passage Resetting Gas}

\author{Marco Biroli\inst{1,2} \and Satya N. Majumdar\inst{2} \and Gr\'egory Schehr\inst{3}}
\shortauthor{M. Biroli \etal}

\institute{                    
  \inst{1} Department of Physics, University of Chicago, Chicago, Illinois 60637, USA\\
  \inst{2} LPTMS, CNRS, Univ.  Paris-Sud,  Universit\'e Paris-Saclay,  91405 Orsay,  France \\
  \inst{3} Sorbonne Universit\'e, Laboratoire de Physique Th\'eorique et Hautes Energies, CNRS UMR 7589, 4 Place Jussieu, 75252 Paris Cedex 05, France
}

\abstract{
We study a one-dimensional gas of $N$ Brownian particles that diffuse independently but are {\it simultaneously} reset whenever {\it any} of them reaches a fixed threshold located at $L > 0$. For any $N > 2$, the system reaches a non-equilibrium stationary state (NESS) at long-times with strong long-range correlations. These correlations emerge purely from the dynamics, and not from built-in interactions. Despite being strongly correlated, the NESS has a solvable structure that allows for an exact computation of several physical observables, both global and local. These include the average density profile, the distribution of the position of the $k$-th ordered particles, the distribution of the 
gap between two consecutive particles and the full counting statistics, i.e., the distribution of the number of particles in a finite interval around the origin.  
}

\begin{document}

\maketitle

Diffusion is a universal model, omnipresent in fields ranging from physics, chemistry, biology, finance to social sciences \cite{C43, VK92, CB14, R10}. In the last decade, stochastic resetting has emerged as an active area of research, with applications in a multitude of areas spanning across disciplines~\cite{EMS20,PKR22}. Stochastic resetting simply means interrupting the natural evolution of a process at random times and restarting from the same initial
condition~\cite{EM11a,EM11b}. There are two main outcomes of this stochastic interruption followed by restarting of a process. First, it can typically drive a system to a nonequilibrium stationary state (NESS). Secondly, in random search processes trying to locate a fixed target, resetting helps to accelerate the search and one can often find an optimal resetting rate that minimises the average time to locate the target \cite{EM11a,EM11b,Reu16,PKE16,PR17,EMS20,B20}. These two aspects have been explored during the last decade in a variety of theoretical models~\cite{EMS20,PKR22,NG23} as well as in experiments on colloidal particles in optical traps~\cite{Roichman20,BBPMC_20,FBPCM_21a,BFPCM_21b,Landauer23,VR25,BCKMPS25}. 

\begin{figure}
    \centering
    \includegraphics[width=\linewidth]{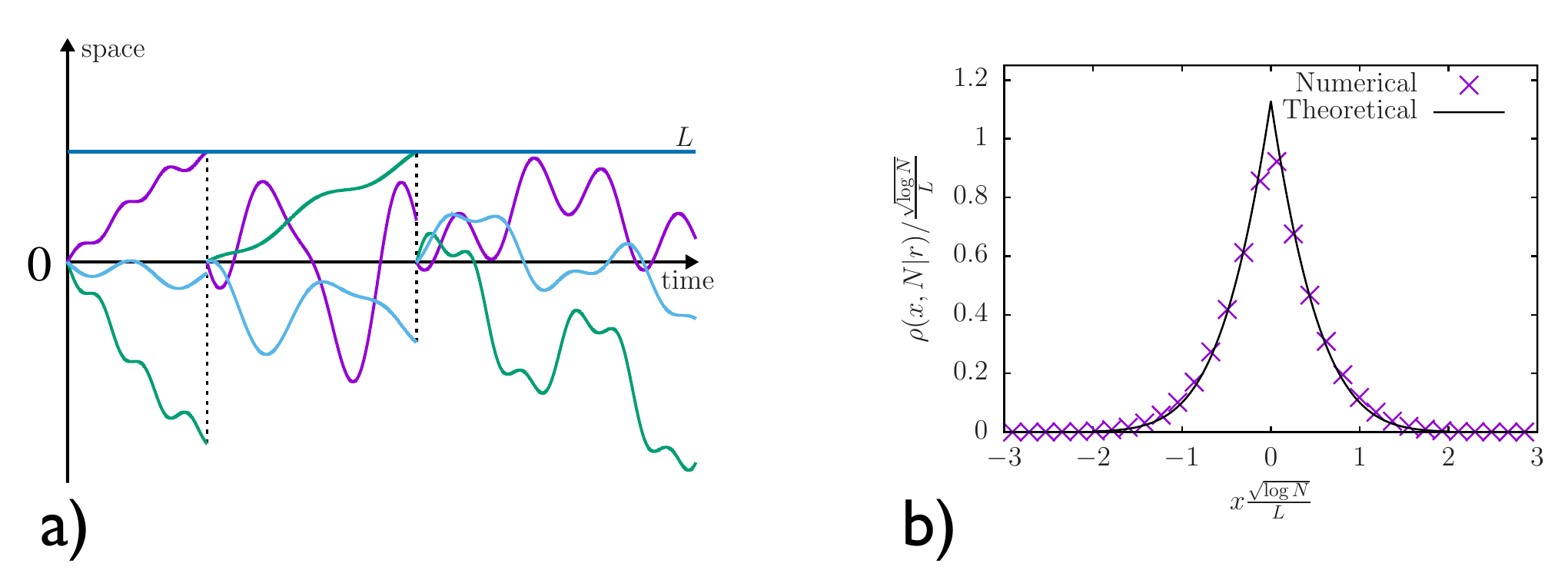}
    \caption{{\bf a):} A schematic representation of the trajectories of three particles (different colors) in the TR model. When any of them reaches $L$, all of them reset to the origin (shown by dashed lines). {\bf b):} The scaling form of the average density $\rho(x,N)$ in Eq. (\ref{eq:f-def}) compared to simulations to $N=10^4$ particles with $D =1$, $L = 1$ and $\dd t = 10^{-5}$, averaged over 100 samples.}
  \label{fig:model}
\end{figure}
While most of the earlier studies focused on the stochastic resetting of a single particle with a variety of dynamics, there have been 
few works on many-body systems. These include fluctuating interfaces, $d$-dimensional Ising model, symmetric exclusion processes on the lattice, etc \cite{BKP19, MMS20,NG23,BMS23,BLMS23, BLMS24,BKMS24,AMG25,MBMS25}. Very recently, a simple model of $N$ noninteracting diffusing particles undergoing {\it simultaneous} stochastic resetting was introduced~\cite{BLMS23}. It was shown that, even though there is no interaction between the particles to start with, the simultaneous resetting introduces all-to-all attractive correlations between the particles that grow with time and eventually saturate to a nonzero value in the stationary state. This is an example of a dynamically emergent correlation (DEC) which comes purely from the dynamics and not from interactions~\cite{BLMS23}. Remarkably, the resulting NESS, though strongly correlated,  has an exactly solvable structure that allows
analytical computations of several observables of this correlated gas. For example, for a one-dimensional resetting gas, 
these observables include the average density (which has a nontrivial profile), the distribution of the rightmost particle (extreme value statistics), the distribution of the spacing between consecutive particles and the full counting statistics (FCS), i.e., the distribution of the number of particles in a certain interval. Such a correlated NESS has since been found for noninteracting particles undergoing non-Brownian dynamics, such as L\'evy flights, and also for non-Poissonian resetting protocols \cite{BLMS24,MBMS25}. Furthermore, a similar solvable NESS with DEC has also been found recently in a variety of theoretical models, both classical \cite{BKMS24,SM2024,MMS25} and quantum \cite{PCML21,MCPL22,KMS25,MKMS25}. These novel DECs have recently been measured also in experiments on optically trapped colloidal particles \cite{VR25,BCKMPS25}. In fact, such emergent correlations have also been found in resetting systems with memory where there is no stationary state~\cite{BM25}.

Most of these studies were performed when the resetting is Poissonian, i.e., occurs with a constant rate and, hence, can be controlled/imposed externally. However, in many situations the resetting can be ``event-driven'', e.g., when a diffusing particle on the line resets to its initial position only when its current position reaches a threshold, say located at $L$. This is generally called ``threshold resetting'' (TR). For a single particle undergoing TR, the position distribution at time $t$ was computed recently in Ref. \cite{BRR20} and it was found that there is no stationary state, at variance with the Poissonian resetting. In a recent study, TR with $N>1$ particles was investigated where all the $N$ particles (Brownian or ballistic) move independently and 
reset simultaneously when the position of one of them hits the threshold $L$~\cite{BMP25}. However, in this work the focus was on the mean first-passage time to find a fixed target in space under TR, and not on the joint position distribution of the particles. The goal of our work is to compute the joint distribution of the positions $x_1, x_2, \cdots, x_N$  exactly in this diffusive TR model at all times $t$. In this model, $N$ Brownian particles, starting at the origin, undergo TR at level $L$ and simultaneously reset to the origin [see Fig. \ref{fig:model}~a)]. We show that this system reaches a nontrivial NESS as $t \to \infty$ for any $N >2$. For $N=1$ and $N=2$, the position distribution remains time-dependent at all times and there is no NESS. Moreover, for $N>2$, we show that this NESS also has DECs and yet has a solvable structure that allows us to compute exactly the statistics of physical observables that were computed before for the Poissonian resetting gas. We will show that these observables exhibit nontrivial scaling behavior in the large $N$ limit, as a result of the event-driven TR mechanism combined with the simultaneous restart of the trajectories.

This simple diffusive TR model of $N$ particles can be useful in a variety of contexts. For example, it can be used to model the behavior of 
systems with a global failure mode, such as $N$ countries' electrical consumption on the same grid with a blackout level, or $N$ stresses along a fault in a stick-slip system with a critical stress level, as in the H\'ebraud-Lequeux model \cite{HL98,B22}. 
%
%
For $N=1$, this model was first introduced and studied by Gerstein and Mandelbrot (GM) in the context of neuronal activity~\cite{GM64}. GM 
pioneered the use of diffusion processes to model neuronal voltage fluctuations under the influence of competing excitatory and inhibitory signals. When the voltage crosses a predetermined threshold, the neuron ``fires", producing a spike, after which the voltage immediately resets to zero. The GM model has served as the basis for numerous works in theoretical neuroscience \cite{SG13, T88}. This is because it is analytically solvable and also serves as a basic toy model for more realistic systems of nerve cell activity \cite{T88}. It also serves as a base for more recent machine learning methods \cite{P25}. However, although experimental data can often be fitted using results from the $N=1$ GM model, it still has limitations \cite{T88}. In their paper \cite{GM64}, GM acknowledged that reducing neuronal dynamics to a single degree of freedom ($N=1$) could be an oversimplification and made initial attempts to extend the model to two interacting particles ($N=2$). For $N>2$, the fate of the system at long times has never been studied and thus remained as an interesting open problem. Our results in this paper provide the first exact results for the stationary state of the GM model with arbitrary $N>2$.  

Our principal finding, namely that the behavior of the system at late times is fundamentally different for $N>2$ and $N \leq 2$, also has interesting implications for the neuronal activity. In firing neural systems one often sees recurrent spikes as a function of time. For the diffusive GM model with $N=1$, such recurrent spikes would be absent since there is no stationary state. In order to model the steady state with recurrent spike patterns, GM actually introduced a drift in the $N=1$ diffusive model. However, experimental data suggest that such spike patterns exist even without a drift~\cite{T88}. The existence of a stationary state in $N>2$ GM model even in the absence of a drift shows that  
a drift is not really necessary for guaranteeing a steady state.

\vspace{0.2cm}


{\bf The model.} We consider $N$ particles diffusing on a line (with the same diffusion constant $D$) with positions denoted by
$x_1(t), \cdots, x_N(t)$. They all start at the origin at $t=0$ and diffuse independently. Whenever any of the coordinates reaches $L$, the 
positions of all $N$ particles are instantaneously reset to the origin. 
%
%
To compute the joint probability distribution function (JPDF) of the positions $P(x_1, \cdots, x_N,t) \equiv P({\bf x},t)$, we need, as basic ingredients, some properties of a single free Brownian motion (without any reset) with an absorbing barrier at $x=L$ and starting at $x=0$. The propagator $G(x,t \vert L)$, denoting the probability density to reach $x$ at time $t$ with an absorbing wall at $L$, can be easily computed using the method of images and it reads~\cite{Redner_book,BMS13,MS24}
\begin{equation} \label{eq:propagator}
    G(x, t \vert L ) = \frac{1}{\sqrt{4 \pi D t}}\left[ e^{-\frac{x^2}{4 D t}} - e^{- \frac{(x - 2L)^2}{4 D t}} \right] \;, \; x \leq L \;.
\end{equation}
Let $S_1(L,t)$ denote the survival probability up to time $t$, i.e., the probability that this walker does not hit the wall 
at $L$ up to time $t$. This can be obtained simply as
\begin{equation} \label{eq:survival-probability}
    S_1(L, t) = \int_{-\infty}^L \dd x \,G(x,t \vert L)\, =  {\rm erf}\left( \frac{L}{\sqrt{4 D t}} \right) \;,
\end{equation}
where ${\rm erf}(z) = (2/\sqrt{\pi}) \int_0^z \dd u\,e^{-u^2}$ is the error function and we used Eq. (\ref{eq:propagator}) for $G(x, t \vert L )$. 
Consequently, the first-passage probability density $F_1(L,t)$ denoting the probability density of crossing $L$ for the first time at time $t$ 
is given by
 \begin{equation} \label{eq:first-passage}
    F_1(L, t) =  -\pdv{S_1(L, t)}{t}= \frac{L}{\sqrt{4 \pi D t^3}} e^{-\frac{L^2}{4 D t}} \;.
\end{equation}
Consider now $N$ independent reset-free particles. Using this independence, 
the survival probability up to time $t$ of this $N$-particle system, i.e., the probability that none of the particles hits $L$
up to time $t$ can be easily computed~as
\begin{equation} \label{eq:SN_to_S1}
    S_N(L, t) = \left[S_1(L, t)\right]^N = \left[{\rm erf}\left( \frac{L}{\sqrt{4 D t}} \right)\right]^N \;.
\end{equation}  
%
%
Finally, the probability density that one of the $N$ particles hits $L$ for the first time at time $t$ follows simply from
\begin{equation} \label{eq:FN_to_F1S1}
    F_N(L, t) =  -\pdv{S_N(L, t)}{t} = N F_1(L, t) \left[S_1(L, t)\right]^{N-1} \;.
\end{equation}
This result can be interpreted very simply. Suppose that the level $L$ is hit for the first time at time $t$. The hitting particle 
can be any one of the $N$ particles and the rest of the $(N-1)$ particles need to survive up to $t$. Given a fixed hitting particle,
the probability of this event is then $F_1(L, t) [S_1(L, t)]^{N-1}$ where we used the independence of the particles. Finally one needs to
multiply this probability by a factor $N$, since the hitting particle can be any one of the $N$ particles. 

We are now ready to consider the $N$ particle system in the presence of TR at $L$. The JPDF $P(\vb{x}, t)$ can be expressed in terms of a renewal equation as
\begin{equation} \label{eq:renewal}
    P(\vb{x}, t) = \prod_{i = 1}^N G(x_i, t \vert L) + \int_0^t \dd t' \; F_N(L, t') P(\vb{x}, t - t') \;,
\end{equation}
where $G(x, t \vert L)$ and $F_N(L, t)$ are given respectively in Eqs.~(\ref{eq:propagator}) and~(\ref{eq:FN_to_F1S1}). This equation can be understood as follows. There are two possibilities for the particles to reach $\vb{x}$ at time $t$. Either they do so in such a way that none of them crosses $L$ up to time $t$ -- this corresponds to the first term where we used the independence of the free particles. The other possibility is that the level $L$ has been reached for the first time at $t' \leq t$ by the reset-free $N$ particle system, which occurs with probability $F_N(L,t') \, \dd t'$. Following this, there is an instantaneous reset to the origin and the process renews itself at time $t'$. Clearly, 
the probability density for this renewed system to reach $\vb{x}$ in time $(t-t')$, starting from the origin, is given by $P(\vb{x}, t - t')$. Finally, we need to integrate $t'$ over $[0,t]$. This explains the second term in Eq. (\ref{eq:renewal}). The convolution structure of Eq. (\ref{eq:renewal}) suggests that it is natural to work in the Laplace space.
Defining the Laplace transform $\tilde{P}(\vb{x}, s) = \int_0^{+\infty} \dd t \; e^{-s t} P(\vb{x}, t)$, we get from Eq. (\ref{eq:renewal}) 
\begin{equation} \label{eq:laplace-renewal-1}
    \tilde{P}(x, s) = \frac{\int_0^{+\infty} \dd t \; e^{-s t} \prod_{i = 1}^N G(x_i,t \vert L)}{1 - \tilde{F}_N(L, s)} \;.
\end{equation}
Taking Laplace transform of the relation $F_N(L,t) = - \partial_t S_N(L,t)$ [see Eq. (\ref{eq:FN_to_F1S1})], and using $S_N(L,0) = 1$ one gets
\begin{equation} \label{eq:laplace-FS-relation}
    \tilde{F}_N(L, s) = 1 - s \, \tilde{S}_N(L, s)  \;,
\end{equation}
which allows us to express Eq. (\ref{eq:laplace-renewal-1}) as
\begin{equation} \label{eq:laplace-renewal-2}
    \tilde{P}(\vb{x}, s) = \frac{1}{s}\,\frac{\int_0^{+\infty} \dd t\; e^{-s t} \, \prod_{i = 1}^N G(x_i, t\vert L)}{\int_0^{+\infty} \dd t\; e^{-s t} \, \left[{\rm erf}\left( \frac{L}{\sqrt{4 D t}} \right)\right]^N} \;, \; x_i \leq L \;.
\end{equation}
Suppose that the system reaches a stationary state at long times, i.e., the JPDF $P(\vb{x},t) \to  P_{\rm st}(\vb{x})$ as $t \to \infty$. Then the Laplace transform  $\tilde{P}(\vb{x}, s) = \int_0^\infty e^{-st} {P}(\vb{x}, t)\, \dd t$ behaves as  $\tilde{P}(\vb{x}, s) \approx P_{\rm st}(\vb{x})/s$ as $s \to 0$. Taking the limit $s \to 0$ in Eq. (\ref{eq:laplace-renewal-2}) we see that it behaves as $\tilde{P}(\vb{x}, s) \approx P_{\rm st}(\vb{x})/s$ where 
\begin{equation} \label{eq:full-jpdf-NESS}
    P_{\rm st}(\vb{x}) = \frac{\int_0^{+\infty} \dd t\; \prod_{i = 1}^N G(x_i,t\vert L)}{\int_0^{+\infty} \dd t' \; \left[{\rm erf}\left( \frac{L}{\sqrt{4 D t'}} \right)\right]^N} \quad, \; x_i \leq L \;, 
\end{equation}
provided, of course, the expression on the right hand side is finite. Indeed, the integral in the numerator is always convergent for any $N \geq 1$. However, the integral in the denominator is convergent only for $N>2$. The denominator just denotes the mean first hitting time of the level $L$ by any of the $N$ walkers and it is finite for any $N>2$. This is because the hitting time distribution has a power-law tail $\sim t^{-N/2-1}$ \cite{MOS11}, whose first-moment is finite only for $N>2$. Thus the JPDF in Eq. (\ref{eq:full-jpdf-NESS}) exists only for $N>2$. In this model, the resetting is event-driven, unlike in the Poissonian resetting case where the resetting rate is constant, which always drives the system to a stationary state \cite{BLMS23}. In our model for $N>2$, since the mean hitting time of the level $L$ is finite, the resetting gets triggered effectively with a constant rate and hence one would expect a stationary state. However, for $N \leq 2$, this resetting event is very rare, implying that it is not strong enough to drive the system to a stationary state. We also notice an interesting and somewhat curious property of the 
stationary JPDF in Eq. (\ref{eq:full-jpdf-NESS}). By rescaling $\tilde t = D\,t$ in Eq. (\ref{eq:full-jpdf-NESS}) -- both in the numerator and in the denominator -- one sees that this stationary JPDF is completely independent of the diffusion constant $D$ for any $N>2$.

One can express the exact stationary JPDF in Eq. (\ref{eq:full-jpdf-NESS}) in a slightly alternative form that explicitly brings out a conditionally independent and identically distributed (CIID) structure as shown below. Using Eq. (\ref{eq:survival-probability}), we note that $\int_{-\infty}^L \dd x\, G(x,t \vert L) = S_1(L,t)$. We then divide and multiply each factor $G(x_i, t \vert L)$ by $S_1(L,t)$ and use (\ref{eq:SN_to_S1}) to rewrite Eq. (\ref{eq:full-jpdf-NESS}) as 
\begin{eqnarray}\label{Pst_CIID}
 P_{\rm st}(\vb{x}) = \frac{\int_0^\infty \dd t\, S_N(L,t)\, \prod_{i=1}^N P_{\rm cond}(x_i \vert t)}{\int_0^\infty \dd t'\, S_N(L,t')}
\end{eqnarray}
where 
\begin{eqnarray} \label{Pcond}
 P_{\rm cond}(x_i \vert t) = \frac{G(x_i,t \vert L)}{S_1(L,t)} \quad, \quad  x_i \leq L \;,
\end{eqnarray}
where, for convenience, we suppressed the $L$-dependence of the left hand side. Clearly, $\int_{-\infty}^L P_{\rm cond}(x_i \vert t) dx_i = 1$. Hence Eq. (\ref{Pst_CIID}) has a probabilistic interpretation: conditioned on the random variable $t$, we have $N$ independent and identically distributed (IID) random variables, each drawn independently from the normalized PDF $P_{\rm cond}(x_i \vert t)$. The random variable $t$ has the distribution
\begin{eqnarray} \label{PDF_t}
q_N(t) = \frac{S_N(L,t)}{\int_0^\infty S_N(L,t')\,\dd t'} \;,
\end{eqnarray}
which is also normalized to unity. A similar CIID structure was found before for $N$ Brownian motions undergoing Poissonian resetting, though there the corresponding conditioning random variable $t$ had just an exponential distribution $q(t) = r\, e^{-rt}$, independently of $N$ \cite{BLMS23}. In fact, in most of the examples of CIID encountered before in the context of stochastic resetting, the distribution of the conditioning distribution was always found to be independent of $N$~\cite{BLMS23,BLMS24,SM2024,MMS25}. Here, for the first time, we find an example where the conditioning distribution $q_N(t)$ depends explicitly on~$N$.

\vspace*{0.5cm}
\noindent
{\bf Large $N$ behavior of $P_{\rm st}(\vb{x})$}. We start from the denominator in Eq. (\ref{Pst_CIID}). The integrand $S_N(L,t) = \left[{\rm erf}\left( \frac{L}{\sqrt{4 D t}} \right)\right]^N$, as a function of $t$, approaches $1$ as $t \to 0$ and vanishes quickly as $t^{-N/2}$ when $t \to \infty$. In the large $N$ limit, the decay is extremely fast as $t$ increases and the integral is dominated by the small $t$ region. In this limit of $N \to \infty$ and $t \to 0$, while keeping the product $t \ln N$ fixed, one can replace the integrand effectively by a sharp step function as 
\begin{equation}\label{SN_approx}
\blue{S_N(L,t)\approx \theta\left(L - \sqrt{4D\,t \ln N}\right)} \;,
\end{equation} 
\blue{where $\theta(z)$ is the Heaviside step function.} With this approximation which is exact in the $N \to \infty$ limit, the denominator (denoted by ${\cal D}$) in Eq. (\ref{Pst_CIID}) becomes, to leading order for large $N$,
\begin{equation}\label{denom_approx}
{\cal D} = \int_0^{+\infty} \dd t \; \left[{\rm erf}\left( \frac{L}{\sqrt{4 D t}} \right)\right]^N \approx \frac{L^2}{4D\,\ln N} \equiv \Lambda \;.
\end{equation} 
Similarly, one can replace $S_N(L,t)$ by the same approximation (\ref{SN_approx}) even in the numerator (denoted by ${\cal N}$) of Eq.~(\ref{Pst_CIID}). This gives, to leading order for large $N$, 
\begin{equation} \label{num_approx}
{\cal N} \approx \int_0^{\Lambda}  \dd t \hspace*{0.cm}\left(\prod_{i=1}^N P_{\rm cond}(x_i \vert t)\right) \;,
\end{equation}
where $P_{\rm cond}(x_i \vert t)$ is given in Eq. (\ref{Pcond}). Furthermore, in the limit $t \to 0$ which dominates the integral in (\ref{num_approx}), one can simplify $P_{\rm cond}(x_i \vert t)$ by noting that $S_1(L,t) = {\rm erf}(L/\sqrt{4Dt}) \approx 1$ and hence 
\begin{equation}\label{num_approx2}
{\cal N} \approx \int_0^{\Lambda}  \dd t \prod_{i=1}^N \frac{1}{\sqrt{4 \pi D t}} \left(e^{-\frac{x_i^2}{4Dt}} - e^{-\frac{(x_i-L)^2}{4Dt}} \right) \;.
\end{equation} 
We now make a change of variable $t =\Lambda/u^2$ and rescale $x_i = \sqrt{4 D \Lambda}\,z_i$. Under this rescaling, the second term in the propagator in Eq. (\ref{num_approx2}) becomes negligible for large $N$ and we get the following scaling behavior for the stationary JPDF 
\begin{equation} \label{eq:jpdf-scaling}
    P_{\rm st}(\vb{x}) \approx \left( \frac{\sqrt{\ln N}}{L} \right)^N \; p_{\rm st}\left( \frac{\sqrt{\ln N}}{L} \vb{x} \right) \;,
\end{equation}
where
\begin{equation} \label{eq:jpdf-NESS-dimensionless}
    p_{\rm st}(\vb{z}) = \int_{1}^{+\infty} \dd u \; \frac{2}{u^3} \prod_{i = 1}^N \frac{u}{\sqrt{\pi}} e^{-u^2 z_i^2 } \;.
\end{equation}
Thus this scaled JPDF $p_{\rm st}(\vb{z})$ also has a CIID structure, namely
\begin{equation} \label{pz_CIID}
p_{\rm st}(\vb{z}) = \int_1^\infty \dd u\, h(u) \, \prod_{i=1}^N p(z_i \vert u) \;,
\end{equation}
where $p(z \vert u ) = u/\sqrt{\pi} e^{-u^2 z^2}$ and $h(u) = 2/u^3$ for $u \in [1,+\infty)$. Thus, conditioned on $u$, the $z_i$'s are IID variables, each drawn from $p(z \vert u)$ and $u$ itself is a random variable drawn from $h(u)$. By making the change of variable $u^2= 1/(2 V)$, we can express this scaling function  in a more suggestive form, namely
\begin{equation} \label{eq:jpdf-NESS-dimensionless2}
 p_{\rm st}(\vb{z}) = 2 \int_0^{1/2} \dd V \,\prod_{i=1}^N \frac{1}{\sqrt{2 \pi V}} e^{-\frac{z_i^2}{2V}} \;.
\end{equation}
Thus this can be interpreted as the joint distribution of $N$ independent Gaussian random variables, each with zero mean and a common variance $V$, with $V$ itself being a random variable distributed uniformly over the interval $[0,1/2]$. This CIID structure in Eq. (\ref{eq:jpdf-NESS-dimensionless2}) clearly shows that the stationary JPDF is not factorizable, indicating that nonzero correlations emerge dynamically from the simultaneous resetting process, even though the walkers are independent, i.e., there is no direct interaction between them. This is thus an example of an event-driven DEC, at variance with models where such correlations emerge due to simultaneous resetting at a constant rate~\cite{BLMS23,BLMS24,SM2024,MMS25}.

To detect this DEC in the stationary state, the natural observable would be the standard connected correlation function $\langle x_i x_j  \rangle - \langle x_i \rangle \langle x_j\rangle$. This however vanishes due to the \blue{$x \to -x$} symmetry of the JPDF in Eq. (\ref{eq:jpdf-scaling}). Hence one needs to consider a higher order correlation measure to reveal these nonzero correlations. The simplest adimensional quantity that does this job is \cite{BM25} 
\begin{equation} \label{def_Ctilde}
\tilde C_{ij} = \frac{\langle x_i^2 x_j^2 \rangle - \langle x_i^2 \rangle \langle x_j^2\rangle}{\langle x_i^4\rangle - (\langle x_i^2 \rangle)^2} \;.
\end{equation}
One can show~\cite{BM25} that $0 \leq \tilde C_{ij} \leq 1$ and takes the minimum value $0$ for uncorrelated variables, while the maximum value $1$ occurs when
they are most strongly correlated, i.e., when $x_i = x_j$. In our case, this correlation $\tilde C_{ij}$ can be easily computed from Eqs. (\ref{eq:jpdf-scaling}) and (\ref{eq:jpdf-NESS-dimensionless2}) and we get, to leading order for large $N$,
\begin{equation} \label{eq:correlator}
   \tilde C_{ij} \approx \frac{1}{33}  \;.
\end{equation}
This correlation does not depend on the indices $i$ and $j$, indicating that any pair of particles has this correlation. Moreover, the fact that $\tilde C_{ij}>0$
indicates that this correlation  between the particles is all-to-all ``attractive''. Physically, such attractive correlations get generated by the simultaneous 
resetting triggered by the threshold reset mechanism in this model. Even though these all-to-all correlations make the stationary state ``strongly correlated'', the
CIID structure of the stationary scaled JPDF in Eq. (\ref{eq:jpdf-NESS-dimensionless2}) allows us to compute several physical observables, both global and local, as we show below.

\noindent {\bf The average density.} Since the system is inhomogeneous, the particles are distributed, globally, with a nontrivial average density profile encoded in the observable $\rho(x,N) = (1/N)\,\left\langle \sum_{i = 1}^N \delta(x - x_i) \right\rangle$ that measures the mean fraction of particles per unit length between $x$ and $x+dx$.
Here $\langle \cdots \rangle$ denotes an average over the stationary JPDF in Eq. (\ref{eq:full-jpdf-NESS}) and is normalized such that $\int_{-\infty}^L dx\, \rho(x,N) = 1$. 
The function $\rho(x)$ can also be expressed as a marginal PDF of a single particle, i.e., 
\begin{equation} \label{eq_rho1}
\rho(x,N) = \int_{-\infty}^L  \dd x_2 \cdots  \int_{-\infty}^L  \dd x_N  P_{\rm st}(x,x_2,\cdots, x_N) \;,
\end{equation}
where $P_{\rm st}({\bf x})$ is given in Eq. (\ref{eq:full-jpdf-NESS}). In the  large $N$ limit, $P_{\rm st}({\bf x})$ takes the scaling form
in Eq. (\ref{eq:jpdf-scaling}). Using this scaling of the JPDF and marginalising, we find that $\rho(x,N)$ satisfies the scaling form
\begin{equation} \label{eq:density-scaling}
    \rho(x,N) \approx \frac{\sqrt{\ln N}}{L} f\left(x \frac{\sqrt{\ln N}}{L} \right) \;,
\end{equation}
where the normalized scaling function $f(z)$ defined for $z \in \mathbb{R}$ is computed by integrating Eq. (\ref{eq:jpdf-NESS-dimensionless}) over $z_2, \cdots z_N$. This gives
\begin{equation} \label{eq:f-def}
    f(z) = \frac{2}{\sqrt{\pi}}\,\int_1^{+\infty}  \frac{\dd u}{u^2}  e^{-u^2 z^2} = \frac{2}{\sqrt{\pi}} e^{-z^2} - 2 |z| \,{\rm erfc}(|z|) \;,
\end{equation}
where ${\rm erfc}(z)$ is the complementary error  function and one can check that $\int_{-\infty}^\infty f(z) \, \dd z = 1$. Note that, although $x \in (-\infty, L]$, the scaled variable $z = x \sqrt{\ln N}/L$ has support over $(-\infty, +\infty)$ in the large $N$ limit. \blue{This scaling demonstrates that the gas of particles is essentially localised over a region of width $\sim 1/\sqrt{\ln N}$ around the origin in the large $N$ limit. This also explains why the scaled JPDF in Eq. (\ref{eq:jpdf-NESS-dimensionless2}) is symmetric in $x$, although in the unscaled coordinates the JPDF in (\ref{Pst_CIID}) is asymmetric due to the presence of the threshold at $x=L$.} The scaling function $f(z)$ is symmetric around $z=0$ with $f(z) \sim 2/\sqrt{\pi} - 2 |z|$ as $z \to 0$ and decays for large $z$ as $f(z) \sim e^{-z^2}/(\sqrt{\pi}\,z^2)$ as $z \to \pm \infty$. This function is plotted in Fig. \ref{fig:model} b) and compared to numerical simulations, showing a perfect agreement. 
\vspace{0.2cm}

\begin{figure}
    \centering
    \includegraphics[width=\linewidth]{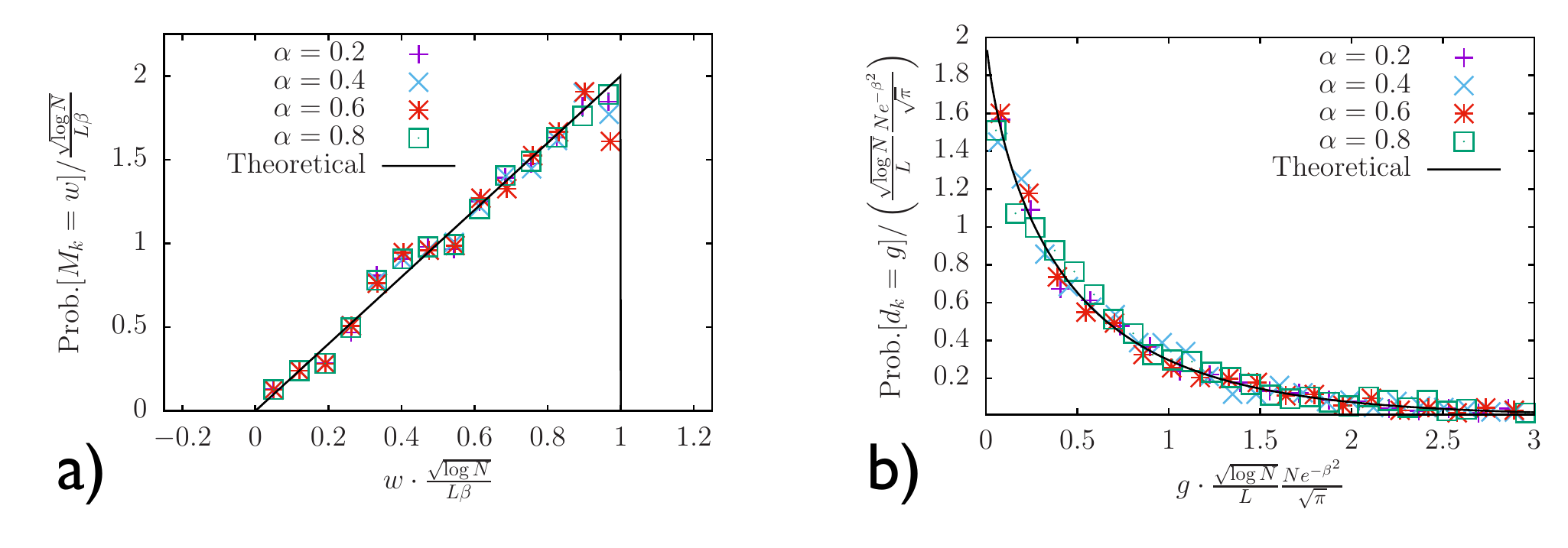}
    \caption{{\bf a)}: The order statistics $M_k$ for $k = \alpha N$ with $\alpha = 0.2, 0.4, 0.6$ and 0.8 in purple (+), blue ($\times$), red ($\star$) and green ($\square$) compared to the theoretical prediction in Eq. (\ref{eq:order}). The numerical points were obtained by sampling directly $10^3$ samples of the distribution defined in Eq. (\ref{eq:jpdf-NESS-dimensionless}) with $N = 10^6$. {\bf b)}: The gap statistics $d_k$ for the same set of $k = \alpha N$ as in {\bf a)} compared to the theoretical prediction in  Eqs. (\ref{eq:h})-(\ref{eq:gap-scaling}). 
    }
    \label{fig:order}
\end{figure}

\noindent{\bf The order statistics.} Now we turn to the order statistics, i.e. the distribution of the ordered positions $M_1 > M_2 > \cdots > M_N$. \blue{In the large $N$ limit,
$P_{\rm st}({\bf x})$ is symmetric under $x \to -x$ and, hence, $M_{k}$ and $- M_{N-k+1}$ have exactly the same statistics.} For instance, for $k=1$, the maximum $M_1$ and the negative of the minimum $-M_N$ have the same statistics. We consider $N$ large and set $k = \alpha N$ where $\alpha \in [0,1]$. Due to the symmetry mentioned above, we need to consider only the regime $\alpha \in [0,1/2)$. It is convenient to consider the scaled ordered maxima $\tilde M_k = M_k \sqrt{\ln N}/L$ such that $\tilde M_k$ remain of $O(1)$ in the large $N$ limit. The distribution of the scaled maxima can be computed from the CIID structure of the JPDF as in (\ref{eq:jpdf-NESS-dimensionless}) using the theory for CIID variables developed in Refs.  \cite{BLMS23,BLMS24}. 
According to this formalism, we first compute the quantile $q(\alpha, u)$ defined as the value above which the average fraction of particle is $\alpha$, where we assume $\alpha \in [0,1/2)$. The quantile is given by   
\begin{equation} \label{eq:quantile-def}
    \alpha = \frac{u}{\sqrt{\pi}}\, \int_{q(\alpha, u)}^{+\infty}  \dd z\,e^{-u^2 z^2}  \;.
\end{equation}
This integral can be computed explicitly and yields
\begin{equation} \label{eq:quantile}
    q(\alpha, u) = \frac{\beta}{u} \quad , \quad {\rm where} \quad  \beta= {\rm erfc}^{-1}(2 \alpha) \;.
\end{equation}
Note that for $\alpha < 1/2$, we have $\beta > 0$ and $ {\rm erfc}^{-1}$ is the inverse of the complementary
error function. From the general theory of CIID variables~ \cite{BLMS23,BLMS24}, it is known that, in the large $N$ limit (for fixed $u$), the distribution  
of the $k$-th maximum $M_k$ can be approximated by \blue{${\rm Prob.}(M_k = w \vert u) \approx \delta\left(w - q(\alpha,u)\right)$}. Averaging this distribution over $u$ drawn from
$h(u) = 2/u^3$ for $u\geq 1$ gives
\begin{equation}
 {\rm Prob.}\left[\tilde M_k = \tilde w\right] \approx \int_1^\infty \dd u \frac{2}{u^3} \delta(\tilde w - q(\alpha,u)) \;.
\end{equation}
Using $q(\alpha,u)$ from Eq. (\ref{eq:quantile}) and performing the integral explicitly one gets $ {\rm Prob.}\left[\tilde M_k = \tilde w\right] = 2 \tilde w/\beta^2$ where $\tilde w \in [0,1]$. Reverting back to the unscaled variable $M_k$, one then finds that its distribution approaches the following scaling form in the large $N$ limit
\begin{equation} \label{eq:order-CIID}
    {\rm Prob.}\left[M_k = w\right] \approx \frac{\sqrt{\ln N}}{L \beta} g\left( w \frac{\sqrt{\ln N}}{L \beta} \right)  \;.
\end{equation}
where the scaling function $g(z)$ is remarkably simple and is given by  
\begin{equation} \label{eq:order}
    g(z) = 2 z \quad {\rm for} \quad 0 \leq z \leq 1 \;.
\end{equation}
Note that it is normalized to unity, i.e., $\int_0^1 g(z) {\rm d}z = 1$. A plot of this scaling form is given in Fig.~\ref{fig:order} a) finding perfect agreement with numerical simulations. Thus, in the large $N$ limit, although the average density is supported over the full space $x \in (-\infty,+\infty)$ as in Eqs. (\ref{eq:density-scaling})-(\ref{eq:f-def}), the typical value of the $k$-th maximum $M_k$ (for $k<N/2$) gets concentrated over a narrow positive interval $M_k \in [0, L \beta/\sqrt{\ln N}]$. For $k>N/2$, the scaling is exactly similar, except that the scaling function is given by $g(z) = -2\,z$ with $z \in [-1,0]$.  

To probe the statistics of the rightmost (or the second, the third, etc) particle, we need to set $k = O(1)$, i.e, $\alpha = O(1/N)$. When $\alpha = c/N$, where $c=O(1)$, one finds that $\beta = {\rm erfc}^{-1}(2\alpha) \sim \sqrt{\ln N}$ and becomes independent of $c$ to leading order for large $N$. Substituting this result in Eq. (\ref{eq:order-CIID}), one finds ${\rm Prob.}[M_k = w] = \frac{1}{L} g\left(\frac{w}{L}\right)$ for any $k=O(1)$, where $g(z)$ is exactly the same scaling function as in Eq. (\ref{eq:order}). Thus the typical values of the first few maxima scale as $O(1)$ in the large $N$ limit, with a linearly increasing concentration over the entire interval $w \in [0,1]$.

\vspace{0.2cm}

\noindent{\bf Gap statistics.} Now we turn to the gap statistics, i.e., the distribution of $d_k = M_k - M_{k+1} > 0$. Here also we set $\alpha<1/2$ without any loss of generality. As before, we consider the scaled gap $\tilde d_k = d_k \sqrt{\ln N}/L$ such that $\tilde d_k$ remains of $O(1)$ in the large $N$ limit. Using again the theory of CIID variables~\cite{BLMS23,BLMS24} as in Eq. (\ref{pz_CIID}) with $h(u) = 2/u^3$ for $u \in [1,+\infty)$, one can show that, conditioned on $u$, the PDF of the scaled gap converges, in the large $N$ limit, to a simple exponential form
\begin{equation} \label{exp_gap}
{\rm Prob.}(\tilde d_k = \tilde g) \approx N\, p(q(\alpha,u)\vert u) \,e^{- N p(q(\alpha,u)\vert u)\, \tilde g} \;,
\end{equation}    
where the conditional distribution $p(z \vert u ) = u/\sqrt{\pi} e^{-u^2 z^2}$ is defined below Eq. (\ref{pz_CIID}) and $q(\alpha,u) = \beta/u$ denotes the quantile defined in Eq. (\ref{eq:quantile}). We now average this exponential distribution (\ref{exp_gap}) over $u \in [1,+\infty)$ drawn from $h(u)=2/u^3$. Performing the integration over $u$ we find that the scaled gap takes the scaling form, ${\rm Prob.}(\tilde d_k = \tilde g) \approx \tilde N \, U\left( \tilde N \, \tilde g \right)$, where $\tilde N = N\,e^{-\beta^2}/\sqrt{\pi}$ with $\beta = {\rm erfc}^{-1}(2\alpha)$ and the scaling function \blue{$U(z)$} is given by  
\begin{equation} \label{eq:h}
    U(z) = 2 z \int_z^\infty \frac{\dd u}{u^2}\,e^{-u} = 2 e^{-z} - 2 z \, \Gamma(0, z) \;,
\end{equation}
where $\Gamma(a, z)$ is the upper-incomplete gamma-function. Reverting back to the original unscaled gap $d_k$, we find that its 
distribution can be expressed in the scaling form
\begin{equation}\label{eq:gap-scaling}
    {\rm Prob.}\left[ d_k = g \right] \approx \frac{\tilde N\,}{L}\, U\left(\frac{\tilde N\, \sqrt{{\log N}} }{L}\, g \right) 
\end{equation}
where the scaling function $U(z)$ is given in Eq. (\ref{eq:h}). It has the asymptotic behaviors 
\begin{equation} \label{eq:h-asymptotics}
    U(z) \approx \begin{cases}
        2 + z (\log z + \gamma - 1) &\mbox{~~when~~} z  \ll 1 \\
        \frac{2 e^{-z}}{z} &\mbox{~~when~~} z \gg 1 \;,
    \end{cases}
\end{equation}
where $\gamma \approx 0.577216$ is Euler's Gamma constant. Numerical simulations, shown in Fig.~\ref{fig:order} b) agree very well with 
the predicted scaling form \blue{in (\ref{eq:gap-scaling})}. 

\vspace*{0.2cm}
\noindent{\bf Full counting statistics (FCS).} We now turn to the FCS, i.e., the distribution of the number of particles $N_{\ell}$ that lie
in the scaled interval $[-\tilde \ell,+ \tilde \ell]$ around the origin where $\tilde \ell = \frac{L}{\sqrt{\log N}} \ell$. We follow exactly the same route as 
for the other observables, namely we exploit the CIID structure in Eq. (\ref{pz_CIID}). 
Skipping details, we find
\begin{equation} \label{eq:fcs-scaling}
    {\rm Prob.}[N_\ell = \kappa  N ] \approx \frac{1}{N} \; H\left( \kappa \right)  \;,
\end{equation}
where the normalized scaling function $H(\kappa)$ is given by
\begin{equation} \label{eq:k}
    H(\kappa) = \frac{\ell^2 \sqrt{\pi}}{{\rm erf}^{-1}(\kappa)^3} e^{{\rm erf}^{-1}(\kappa)^2} \;, \; \kappa_{\min} \leq \kappa \leq 1
\end{equation}
defined on $\kappa \in [\kappa_{\min}, 1]$ with $\kappa_{\min} = {\rm erf}(\ell)$. The asymptotics of $H(\kappa)$ are given by
\begin{equation}
    \label{eq:G-asymptotic}
    H(\kappa) \approx \begin{cases}
        \frac{e^{\ell^2} \sqrt{\pi}}{\ell} + (\kappa - \kappa_{\min}) \frac{e^{2 \ell^2} (2 \ell^2 - 3) \pi}{2 \ell^2}, &\mbox{~for~} \kappa \to \kappa_{\min} \\
        \frac{\ell^2 }{\log({1 - \kappa})^{2} (1 - \kappa)}, &\mbox{~for~} \kappa \to 1 \;.
    \end{cases} 
\end{equation}
We have checked that numerical simulations agree well with our theoretical predictions, as in the case of other observables. The existence of a finite $\kappa_{\min} > 0$ thus points to the following physical picture. This 
means indeed that the total number of particles inside the interval $[-\tilde \ell, + \tilde \ell]$ must be at least
$\kappa_{\min}\, N$ in the large $N$ limit. Thus it is very unlikely to have a rarefied region around the origin where $N_\ell$ is small. 
This happens due to the fact that, under repeated threshold resettings, the particles get squeezed
in a narrow region of size $L/\sqrt{\ln N}$ around the center. Thus, in a typical configuration, there is always a large number of particles (at least $O(\kappa_{\min}\, N)$) around the origin.  

To summarize, we have shown that when the simultaneous resetting of $N>2$ noninteracting particles is triggered by the breaching of a barrier/threshold at $L$, it correlates the particles and drives them into a nontrivial NESS at late times. Even though the particles are strongly correlated in this NESS, the stationary joint distribution has a CIID structure that enables us to compute exactly the distributions of several physical observables in the large $N$ limit and they exhibit nontrivial
scaling behaviors, which agree very well with numerical simulations. Unlike in the standard resetting protocols used before, here the effective resetting rate depends 
on the number of particles $N$. 

One can envisage several applications of our results, since this is a natural generalization to $N>2$ case of the classical fire and integrate 
model of neuron dynamics ($N=1$) in Ref.~\cite{GM64}. Our results show that, even if the neurons are noninteracting, they can get
strongly correlated via repeated firing followed by resetting events. In this basic model there is no direct interaction between neurons and it would be interesting to investigate the stationary state in the presence of both interaction and the DECs.  
{\acknowledgments} We would like to thank V. Hakim, G. L. Pinter, S.~Ostojic and S. Waxman for useful discussions. We acknowledge support from ANR Grant No. ANR-23-CE30-0020-01 EDIPS. This research received support through Schmidt Sciences, LLC.




\end{document}